\documentclass[preprint,aps]{revtex4}

\usepackage{graphicx}
\usepackage{dcolumn}
\usepackage{bm}
\usepackage{color}
\usepackage{amsmath}

\begin{document}
\title{Spatial control of surface plasmon polariton excitation at planar metal surface}
\author{Zhichao Ruan$^{1,2}$, Hui Wu$^{3}$, Min Qiu$^{2,4}$, and Shanhui Fan$^{5}$ }
\affiliation{$^1$ Department of Physics, Zhejiang University, Hangzhou 310027, China \\
$^2$ State Key Laboratory of Modern Optical Instrumentation, Zhejiang University, Hangzhou 310027, China \\
$^3$ State Key Laboratory of New Ceramics and Fine Processing, School of Materials Science and Engineering, Tsinghua University, Beijing 100084, China  \\
$^4$ Department of Optical Engineering, Zhejiang University, Hangzhou 310027, China \\
$^5$ Ginzton Laboratory, Department of Electrical Engineering, Stanford University, Stanford, California 94305}

\begin{abstract}
We illustrate that the surface plasmon polariton (SPP) excitation through the prism coupling method is fundamentally limited by destructive interference of spatial light components. We propose that the destructive interference can be canceled out by tailoring the relative phase for the different spatial components. As a numerical demonstration, we show that through the phase modulation the excited SPP field is concentrated to a hot energy spot, and the SPP field intensity is dramatically enhanced about three folds in comparison with a conventional Gaussian beam illumination.
\end{abstract}

\maketitle
Efficiently exciting and focusing surface plasmon waves at a given site are important for nanophotonic applications. Recently, particular attention has been attracted by using the spatial control techniques to control the excitation and concentration of light in a variety of optical systems. Vellekoop {\it{et.~al.}} demonstrated that sending light through a random scattering medium, with a spatially modulated input wave front, can achieve the constructive interference at a focus spot, resulting in focusing beyond the diffraction limit \cite{vellekoop2010exploiting}. This effect can be interpreted as time-reversal of light emitted from a point emitter  \cite{derode1995robust}. Similar idea has also been applied to generate and control the location of light hot spots in a diffraction grating \cite{PhysRevLett.101.013901} and to experimentally focus surface plasmon wave on a metal surface of a nanohole array \cite{gjonaj2011active}. In another related work, Kao {\it{et.~al.}} proposed that in a plasmonic metamaterial the strong optically induced interactions between discrete metamolecules can be manipulated to realize a subwavelength scale energy localization through spatially tailoring the phase profile of a continuous-wave input light beam \cite{Zheludev2011coherent}. Conceptually, these spatial control methods have the strong analogy to the coherent control methods where an incident optical pulse is modulated in the time domain to steer light-matter interaction \cite{rabitz2000whither, weinacht1999controlling,Silberberg2002coherent,ginsberg2007coherent} or the response of optical systems \cite{Stockman2002coherent,Stockman2004coherent,aeschlimann2007adaptive,aeschlimann2010spatiotemporal, sandhu2010enhancing} towards a desired final state.

Since a flat metal surface is the simplest geometry sustaining surface plasmons, it is of fundamental importance in plasmonics to efficiently excite the SPP field on the surface \cite{maier2007plasmonics}. The established excitation techniques, including the prism coupling method, have practical significance for applications including surface-enhanced sensing and spectroscopy \cite{homola2008surface,liu2011long}, plasmonic nonlinear optics\cite{palomba2008nonlinear,renger2010surface,grosse2012nonlinear}, and plasmon optical tweezers \cite{volpe2006surface,righini2007parallel,juan2011plasmon}. Here we propose that the SPP excitation on a flat metal surface can be dramatically enhanced by phase modulation of each wave-vector component of an incident light. We show that under a conventional illumination the spatial components of the incident light have destructive interference on the SPP excitation, and it strongly limits the SPP excitation and lowers the field intensity. We demonstrate that the destructive interference can be canceled out by shaping illumination beam with a tailored relative phase for the different wave-vector components. As a consequence, the excited SPP field under the phase-shaped beam illumination is concentrated to a hot energy spot, and the electric field intensity is enhanced about three folds at the peak in comparison with the conventional Gaussian beam illumination. The proposed phase-shaped beam approach provides a new degree of freedom to fundamentally control the SPP excitation.

\begin{figure*}
\centerline{\includegraphics[width=6.5in]{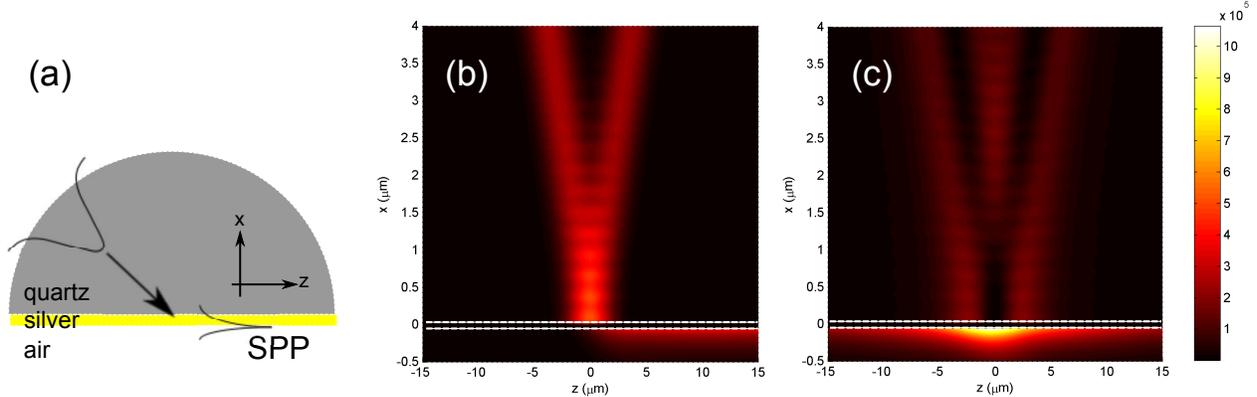}}
\caption{\label{fig:schematic} (a) Schematic of a SPP excitation with the Kretschmann configuration. The thickness of the silver layer is $58.3$nm, and the dielectric constant of the quartz prism is ${\varepsilon _d} = 2.25$. (b-c) Distribution of the electric field intensity ${\left| {\bf{E}} \right|^2}$ under the illumination of a Gaussian beam and the phase-shaped beam with the phase modulation of Eq~(\ref{eq:phase_modulation}). Here the white dashed lines outline the interfaces of the metal layer.}
\end{figure*}

To clearly show the interference effect, we develop a spatial coupled-mode formula to model the SPP excitation process. To excite the SPP field at a metal-dielectric surface the prism coupling method of the Kretschmann and Otto configurations is commonly employed to satisfy the phase-matching requirement. Our theory can be generally applied to the prism coupling cases. As an example, here we consider a Kretschmann configuration [Fig.~\ref{fig:schematic}(a)], where a $p$-polarized incident beam illuminates a metal layer coating on a quartz prism. When the parallel component of the incident wave vector, $k_z$, is close to the SPP wave vector, the incident light excites SPP through the evanescent wave. Meanwhile the excited SPP leaks out and generates the radiation wave in quartz as it propagates along the $z$ direction. Therefore the reflection process consists of two pathways: the direct reflectance of the incident wave at the quartz-metal interface, and the outgoing radiation from the leakage of the excited SPP at the metal-air interface.

Based on the spatial coupled-mode theory (CMT) formalism \cite{haus1984waves}, such a excitation process can be described by the following equations:
 \begin{subequations}
 \label{eq:coupled}
\begin{eqnarray}
 \frac{{da}}{{dz}} &=& (i{\beta _{spp}} - {\alpha _{l}} - {\alpha _{spp}})a + i{e^{i\phi }}\sqrt {2{\alpha _{l}}} {S_{in}}(z) \label{eq:coupled_mode_1}\\
 {S_{out}} &=& {e^{i\phi }}{S_{in}} + i{e^{i\phi }}\sqrt {2{\alpha _{l}}} a \label{eq:coupled_mode_2}
 \end{eqnarray}
 \end{subequations}
Here we take the convention that the field varies in time as $\exp(-i\omega t)$. $a$ is the amplitude of the SPP which is normalized such that ${\left| a \right|^2}$ corresponds to the time-averaged power along $z$ direction. $S_{in}(z)$ ( $S_{out}(z)$ ) corresponds to the amplitude of the incident (reflective) light with the normalization of ${\left| S_{in} \right|^2}$ (${\left| S_{out} \right|^2}$ ) giving the $x$-direction Poyntin flux.  $\phi$ is the phase change of the reflection at the quartz-metal interface. $\alpha_{l}$ and $\alpha_{spp}$ correspond to the loss rate of the SPP due to the leaky radiation and that of the propagation loss due to intrinsic material absorption, respectively. Note that the spatial CMT takes the approximation of the strong confinement condition, i.e. $\alpha_{l}+\alpha_{spp}<<\beta_{spp}$, and the direct reflection coefficient at the quartz-metal surface is the same for all the parallel wave vectors around $\beta_{spp}$ in both lossless and lossy cases. We will see below in the numerical example that such an approximation is in fact sufficient for the SPP excitation at a real silver interface.

Let us consider two different scenarios with the spatial CMT theory. First, when the metal has an infinite thickness, the leaky rate $\alpha_{l}=0$. In this case, Eqs.(\ref{eq:coupled_mode_1}) and (\ref{eq:coupled_mode_2}) are decoupled: The stable solution $a = \exp (i{\beta _{spp}}z - {\alpha _{spp}}z)$ of Eq.(\ref{eq:coupled_mode_1}) presents the SPP propagating in the $z$ direction, and Eq.(\ref{eq:coupled_mode_2}) turns out as ${S_{out}} = {e^{i\phi }}{S_{in}}$, i.e. the reflected light solely experiences the phase change at the quartz-metal surface.  In contrast, if the metal thickness is finite, Eq.(\ref{eq:coupled_mode_1}) shows the SPP excitation through the mode coupling. Eq.(\ref{eq:coupled_mode_2}) indicates the reflective light contributed from the SPP excitation and the direct reflection at the quartz-metal surface. In this case, we expand the incident (reflected) beam into a series of plane waves as ${S_{in(out)}} = \int_{ - \infty }^\infty  {{s_{in(out)}}({k_z})\exp (i{k_z}z)d{k_z}}$. On substituting into Eq.(\ref{eq:coupled}), the amplitude of the excited SPP and the reflection coefficient are obtained
\begin{subequations}
 \label{eq:coupled2}
\begin{eqnarray}
 a(z) &=& \sqrt {2{\alpha _{l}}} {e^{i\phi }}\int\limits_{ - \infty }^\infty  {\frac{{{s_{in}}({k_z})\exp (i{k_z}z)}}{{({k_z} - {\beta _{spp}}) - i({\alpha _l} + {\alpha _{spp}})}}d{k_z}}  \label{eq:amp_SPP}\\
 R & \equiv & \frac{{{s_{out}}}}{{{s_{in}}}} = {e^{i\phi }}\frac{{({k_z} - {\beta _{spp}}) + i{\alpha _l} - i{\alpha _{spp}}}}{{({k_z} - {\beta _{spp}}) - i{\alpha _l} - i{\alpha _{spp}}}}  \label{eq:reflection_SPP}
 \end{eqnarray}
 \end{subequations}

Eq.(\ref{eq:amp_SPP}) indicates that the amplitude of the excited SPP is sensitive to the phase of the incident spatial components $s_{in}(k_z)$. Particularly, the real part of the integrand's denominator changes sign about $\beta_{spp}$. Therefore, when the incident light is a conventional beam that $s_{in}(k_z)$ has linear phase variation about $\beta_{spp}$, such as a Gaussian illumination, the components with $k_z<\beta_{spp}$ and $k_z>\beta_{spp}$ have opposite contributions to the SPP excitation. As a result, a destructive interference arises, which strongly limits the peak intensity of the excited field. The analogous destructive interference also occurs in the time domain, where a excited state is subject to a conventional un-chirped pulse in the applications of optically induced resonant transitions \cite{Silberberg2002coherent} and all-optical bistable switching \cite{sandhu2010enhancing}. Thus, in order to cancel out such destructive interference, a wavevector-dependent phase modulation on $s_{in}$ is required to balance the phase change around the SPP resonance:
\begin{equation}  
{\tilde s_{in}} = {s_{in}(k_z)}\exp (i\arg [({k_z} - {\beta _{spp}}) - i({\alpha _l} + {\alpha _{spp}})] )
\label{eq:phase_modulation}
\end{equation}
This phase modulation can be achieved experimentally using commercially available spatial light modulators (SLM).

\begin{figure}
\centerline{\includegraphics[width=3.2in]{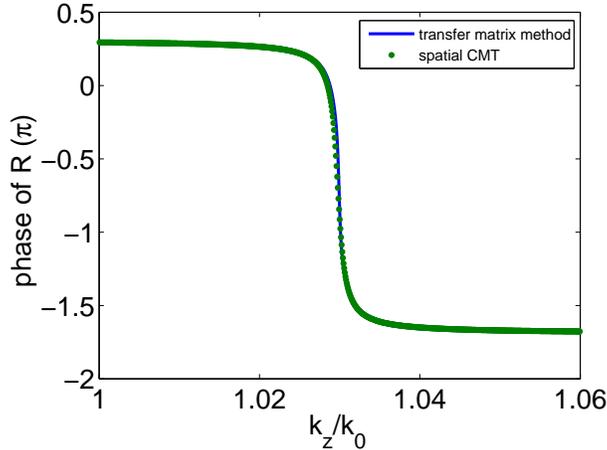}}
\caption{\label{fig:phase_lossless} Phase spectrum of the reflection coefficient in the lossless case. The solid line corresponds to the numerical calculation by the transfer matrix method. The dotted line is the fitting result of the CMT Eq.(\ref{eq:reflection_SPP}) with the parameters:  $\phi=0.97009$, $\beta_{spp}= 1.0298k_0$, $\alpha_{spp}=0$, and $\alpha_l=6.6990 \times {10^{ - 4}}k_0$.}
\end{figure}

We numerically demonstrate this idea of enhancing the SPP excitation by a spatially-modulated beam in a Kretschmann configuration (Fig.~\ref{fig:schematic}), where the thickness of the metal layer is $58.3$nm, and the dielectric constant of the quartz prism is ${\varepsilon _d} = 2.25$. We first neglect the material absorption and assume the metal with a Drude permittivity dispersion ${\varepsilon _m} = {\varepsilon _\infty } - {\omega _p^2}/({{\omega ^2} + i{\gamma _d}\omega })$, $\varepsilon _\infty=4.039$, $\omega_p=1.391 \times {10^{16}}/(2\pi)$Hz, $\gamma_d=0$. In this lossless case, the propagation loss rate $\alpha_{spp}=0$, and as indicated by Eq.(\ref{eq:reflection_SPP}) the amplitude of $R$ is unity agreeing with energy conservation.

To determine the other parameters of the spatial CMT theory, we fit Eq.(\ref{eq:reflection_SPP}) to the phase of the reflection coefficient obtained from numerical simulation. Assuming that the wavelength of the incident light in free-space is $\lambda_0=632$nm, the solid line in Fig.~\ref{fig:phase_lossless} corresponds to the phase spectrum of the reflection coefficient calculated by the transfer matrix method \cite{chew1995waves}. From Eq.(\ref{eq:reflection_SPP}) we see that the phase of $R$ should have a $2\pi$ shift as $k_z$ changes from $k_z \ll \beta_{spp}$ to $k_z \gg \beta_{spp}$. In the range of $k_{z}=1 \sim 1.06k_0$ ($k_0=2\pi/\lambda_0$), the numerical simulation indicates that the phase of $R$ exhibits such a $2\pi$ shift. The theoretical results from Eq. (2), using the fitting  parameters of $\phi=0.97009$, $\beta_{spp}= 1.0298k_0$, and $\alpha_l=6.6990 \times {10^{ - 4}}k_0$, are plotted as the dotted line in Fig.~\ref{fig:phase_lossless}, which agree well with the numerical calculation. Also, $\beta_{spp}$ as determined from the fit to the numerical simulation coincides with the wave vector of non-leaky SPP at the metal-air interface, i.e. ${k_0}\sqrt {{\varepsilon _m}/(1 + {\varepsilon _m})}$, and the very small deviation is due to the leakage of the SPP field.

\begin{figure}
\centerline{\includegraphics[width=3.5in]{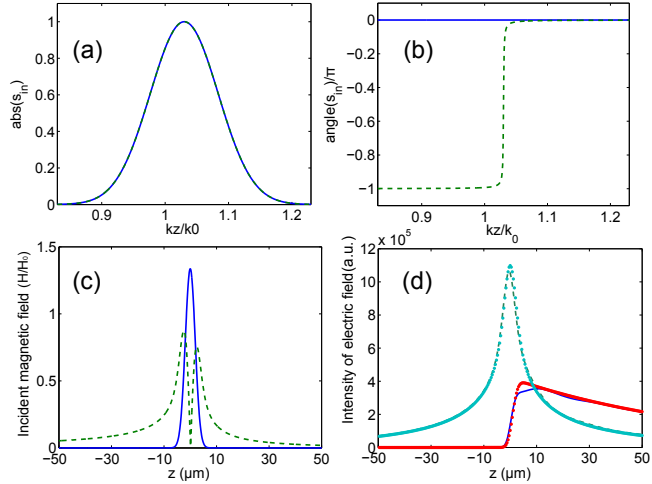}}
\caption{\label{fig:phase_modulation} (a-b) The amplitude and phase spectra of the spatial components for the Gaussian beam (solid) and the phase-shaped beam Eq.~(\ref{eq:phase_modulation}) (dashed), respectively. (c) The magnetic field amplitude of the Gaussian beam (solid) and the phase-shaped beam (dashed) at the metal-quartz interface. (d) The electric intensity of the excitation SPP. The blue solid and green dashed lines correspond to Figs.~\ref{fig:schematic}(b-c) at $x=-0.06\mu$m, respectively. The red and cyan dotted lines are computed by the CMT Eq.~(\ref{eq:amp_SPP}) with the fitting parameters.}
\end{figure}

The enhancement of SPP excitation is illustrated by a comparison between a conventional Gaussian illumination and the shaped beam after the phase modulation. We consider that the incident Gaussian beam has the magnetic field ${H_y} = {H_0} \int {s_{in}({k_z})} \exp (i{k_z}z + i{k_x}x)d{k_z}$, where $H_0$ is a normalization constant, $s_{in}({k_z}) = \exp [ - ({k_z} - {\beta _{spp}})^2/k_w^2]$, and $k_w = 0.0758k_0$. Such a Gaussian beam has a $2.652\mu$m radius waist and focuses at the quartz-metal surface, and the incident angle is $43.4^{\circ}$. Figs.~\ref{fig:phase_modulation}(a-b) are the amplitude and phase spectra for the Gaussian beam and the phase-shaped beam with the phase modulation of [Eq.~(\ref{eq:phase_modulation})]. Fig.~\ref{fig:phase_modulation}(c) shows the incident magnetic field amplitude at the quartz-metal surface. Distinct from the Gaussian beam having a peak at the center, the magnetic field of the phase-shaped beam is zero at the center. Also the phase-shaped beam is much wider than the Gaussian beam and does not exhibit a symmetry profile about the center.

\begin{figure}
\centerline{\includegraphics[width=3.0in]{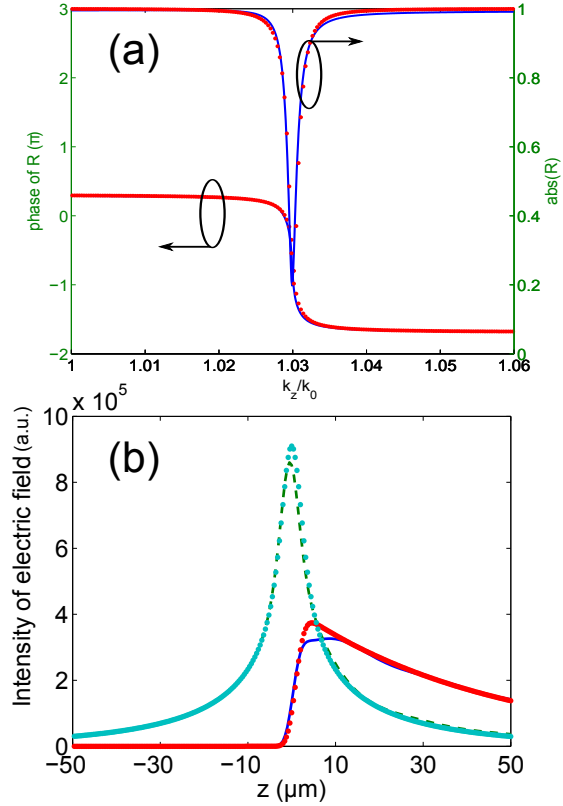}}
\caption{\label{fig:lossly_coherent}  (a) The amplitude and phase spectra of the reflection coefficient in the lossy case of the real silver. The solid line corresponds to the numerical calculation. The dotted line is the fitting result of the CMT Eq.(\ref{eq:reflection_SPP}) with the parameters:  $\phi=0.97009$, $\beta_{spp}=1.0301k_0$,  $\alpha_{spp}=4.5062 \times {10^{ - 4}}k_0$, and $\alpha_l=6.6990 \times {10^{ - 4}}k_0$. (b) The electric intensity of the excitation SPP at $x=-0.06\mu$m. The blue solid and green dashed lines respectively correspond to the cases of the Gaussian beam and the phase-shaped beam, calculated by the numerical simulation. The red and cyan dotted lines are computed by the CMT Eq.~(\ref{eq:amp_SPP}) with the fitting parameters.}
\end{figure}

Figs.~\ref{fig:schematic}(b-c) show the electric field intensity under the Gaussian beam and the phase-shaped beam illuminations. The fields are computed by using the superposition of the different spatial components applying the transfer matrix method to each spatial component. In comparison with the Gaussian beam, and the phase-shaped beam illumination the incident light efficiently tunnels through the metal layer and excites a SPP spot with much stronger intensity than that of the Gaussian beam. The intensities of the excited SPP at the surface $x=-0.06\mu$m are plotted as the blue and green solid lines in Fig.~\ref{fig:phase_modulation}(d) for the Gaussian beam and the phase-shaped beam illuminations, respectively. To compare with our spatial CMT, the red and cyan dotted lines in Fig.~\ref{fig:phase_modulation}(d) correspond to the intensity of excited SPP calculated by Eq.~(\ref{eq:amp_SPP}) with the fitting parameters and the normalization process, which show good agreement between the coupled-mode theory and the numerical simulation. As our theory predicts, the SPP field excited with the phase-shaped beam has a much higher peak intensity, and it is indeed more than three-fold stronger than the conventional Gaussian illumination case. We note that the phase-shaped beam has the same incident power as the conventional Gaussian beam. More interestingly, the excited SPP field under the phase-shaped beam illumination is concentrated, and the hot energy spot has the full-width half-maximal about $10\mu$m and exhibits a symmetry profile about $z=0$.

We now take into account the material absorption of metal by setting the damping rate ${\gamma _d=}3.139 \times {10^{13}/(2\pi)}$Hz, based on experimentally retrieved silver dispersion at the room temperature \cite{Johnson1972Optical}. With the numerical simulation, the amplitude and phase spectra of the reflection coefficients are calculated and shown as the solid lines in Fig.~\ref{fig:lossly_coherent}(a). We fit the spectra with Eq.~(\ref{eq:reflection_SPP}) using the parameters $\phi=0.97009$, $\beta_{spp}=1.0301k_0$,  $\alpha_{spp}=4.5062 \times {10^{ - 4}}k_0$, and $\alpha_l=6.6990 \times {10^{ - 4}}k_0$. The fitting results are plotted as the dotted lines in Fig.~\ref{fig:lossly_coherent}(a), which indicates good agreement between theory and simulation. Using both the numerical simulation and the spatial CMT Eq.~(\ref{eq:amp_SPP}) with the fitting parameters, the intensity of electric field under the Gaussian beam and the phase-shaped beam illumination are calculated and plotted as the blue and green solid lines in Fig.~\ref{fig:lossly_coherent}(b). Again, the calculation through the CMT theory agrees well with the numerical simulation. Both the simulation and the theory show that in the real lossy metal case the intensity of the excited SPP with the phase-shaped beam is still 2.7 folds stronger than that of the conventional Gaussian illumination.

\begin{figure}
\centerline{\includegraphics[width=3.5in]{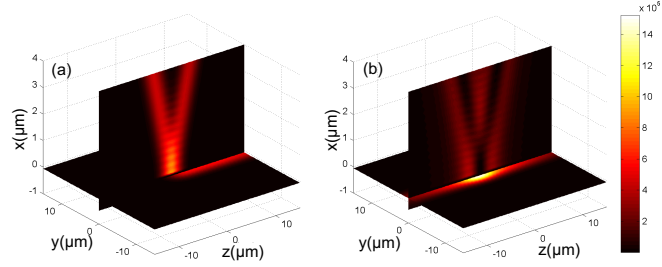}}
\caption{\label{fig:3d}  Distribution of the electric field intensity ${\left| {\bf{E}} \right|^2}$ under the illumination of (a) the 3D $p$-polarization Gaussian beam and (b) the phase-shaped beam with the phase modulation of Eq~(\ref{eq:phase_modulation}). The Gaussian beam has a $2.652\mu$m radius waist and focuses at the quartz-metal surface.}
\end{figure}

The proposed spatial control technology can be directly applied to three dimensional (3D) beams which have a confined profile in the $y$-direction. In these cases, Eq.~(\ref{eq:phase_modulation}) need take into account the wave-vector components in both $y$ and $z$ directions. Fig.~\ref{fig:3d}(a) and (b) present the simulation results for the 3D $p$-polarized Gaussian beam and the phase-shaped beam through the modulation as Eq.~(\ref{eq:phase_modulation}), respectively. The excited SPP field by the phase-shaped beam still concentrates as a hot spot, and the peak intensity is $2.7$ times the Gaussian beam illumination.

In summary, we propose that the SPP excitation on a metal surface can be strongly enhanced by the phase modulation on the incident illumination. We note that the field enhancement results from the constructive interference between different wavevector components, while the launching efficiency for the SPP mode of $k_z=\beta_{spp}$ is the same for both the conventional illumination and the phase-shaped beam. The proposed spatial control technique benefits the development of surface-enhanced applications. For example, in plasmon optical tweezers, a very high and sharp field intensity of SPP excitation leads to strong optical gradient forces. Moreover, the phase shaping provides a new degree of freedom to fundamentally control the SPP excitation to move or sort micro-objects, and the spatial coupled-mode theory is vital to design the phase modulation for a demanded intensity profile.

This work was financially supported by Fundamental Research Funds for the Central Universities (2014QNA3007).  H. W. acknowledges the supports from National Basic Research of China (Grant No. 2013CB632702) and NSF of China (Grant No. 51302141).


\end{document}